\documentclass[runningheads]{svmult}

\usepackage{makeidx}   
\usepackage{graphicx}  
\usepackage{subeqnar}  
\usepackage{multicol}  
\usepackage{physprbb}  
\makeindex             

\def\rn{\noindent\parshape 2 0truecm 8.5truecm 0.3truecm 8.2truecm}
\def\rn{}
\def\nn#1 #2{#2. #1}				
\def\nnn#1 #2 #3{#2. #3. #1}			
\def\nnnn#1 #2 #3 #4{#2. #3. #4 #1}		
\def\nnnnn#1 #2 #3 #4 #5{#2. #3. #4 #5. #1}	
\def\dualand{ and\hbox{ }}				
\def\multiand{, and\hbox{ }}				
\def\rf#1;#2;#3;#4;#5 {{\frenchspacing\par\rn#1: #3 {\bf #4}, #5 (#2). \par}}
\def\rfbook#1;#2;#3;#4;#5 {{\frenchspacing\par\rn#1: {\it #3} (#5, #4, #2).\par}}
\def\rfprep#1;#2;#3 {{\par\frenchspacing\rn#1: #3 (#2).\par}}

\def\bredd{0.8}	

\def\meter{{\rm m}}
\def\kg{{\rm kg}}

\def\Mpc{{\rm Mpc}}

\def\eV{{\rm eV}}

\def\etal{{\frenchspacing\it et al.}}
\def\ie{{\frenchspacing\it i.e.}}
\def\eg{{\frenchspacing\it e.g.}}

\def\beq#1{\begin{equation}\label{#1}}
\def\eeq{\end{equation}}
\def\beqa#1{\begin{eqnarray}\label{#1}}
\def\eeqa{\end{eqnarray}}

\def\fig#1{Figure~\ref{#1}}

\def\spose#1{\hbox to 0pt{#1\hss}}
\def\simlt{\mathrel{\spose{\lower 3pt\hbox{$\mathchar"218$}}
     \raise 2.0pt\hbox{$\mathchar"13C$}}}
\def\simgt{\mathrel{\spose{\lower 3pt\hbox{$\mathchar"218$}}
     \raise 2.0pt\hbox{$\mathchar"13E$}}}
\def\simpropto{\mathrel{\spose{\lower 3pt\hbox{$\mathchar"218$}}
     \raise 2.0pt\hbox{$\propto$}}}

\def\Ob{\Omega_{\rm b}}
\def\Oc{\Omega_{\rm cdm}}
\def\Od{\Omega_{\rm dm}}
\def\Ok{\Omega_{\rm k}}
\def\Ol{\Omega_\Lambda}
\def\Om{\Omega_{\rm m}}
\def\On{\Omega_\nu}
\def\ob{\omega_{\rm b}}

\def\od{\omega_{\rm dm}}

\def\on{\omega_\nu}
\def\fn{f_\nu}

\def\ns{n_s}
\def\nt{n_t}
\def\As{A_s}
\def\At{A_t}

\def\zion{z_{ion}}

\def\p{{\bf p}}

\def\I{{\bf I}}

\def\l{\ell}

\begin{document}
\title*{Latest cosmological constraints on the 
densities of hot and cold dark matter}
\toctitle{Latest cosmological constraints on the 
densities of hot and cold dark matter}
\titlerunning{Cosmological constraints on dark matter}
%
\author{Max Tegmark\inst{1}
\and Matias Zaldarriaga\inst{2}
\and Andrew~J.~S.~Hamilton\inst{3}}
\authorrunning{Max Tegmark et al.}
%

\institute{Dept. of Physics, Univ. of Pennsylvania, 
Philadelphia, PA 19104; max@physics.upenn.edu
\and 
Institute for Advanced Study, Princeton, 
NJ 08540; matiasz@ias.edu
\and
JILA and Astrophysical and Planetary Sciences,
Box 440, Univ. of Colorado,
Boulder, CO 80309; Andrew.Hamilton@colorado.edu
}

\maketitle              

\begin{abstract}

As experimentalists step up their pursuit of 
cold dark matter particles and neutrino masses, 
cosmological constraints are tightening.
We compute the joint constraints on 11 cosmological parameters
from cosmic microwave background and large scale structure data,
and find that at 95\% confidence,
the total (cold+hot) dark matter density 
is $h^2\Od=0.20^{+.12}_{-.10}$
with at most 38\% of it being hot (due to neutrinos). 
A few assumptions, including negligible neutrinos,
tighten this measurement to
$h^2\Od=0.13^{+0.04}_{-0.02}$, \ie, 
$2.4\times 10^{-27}\kg/\meter^3$ give or take 20\%.
\end{abstract}

\begin{figure}[tb]
\begin{center}
\includegraphics[width=\bredd\textwidth]{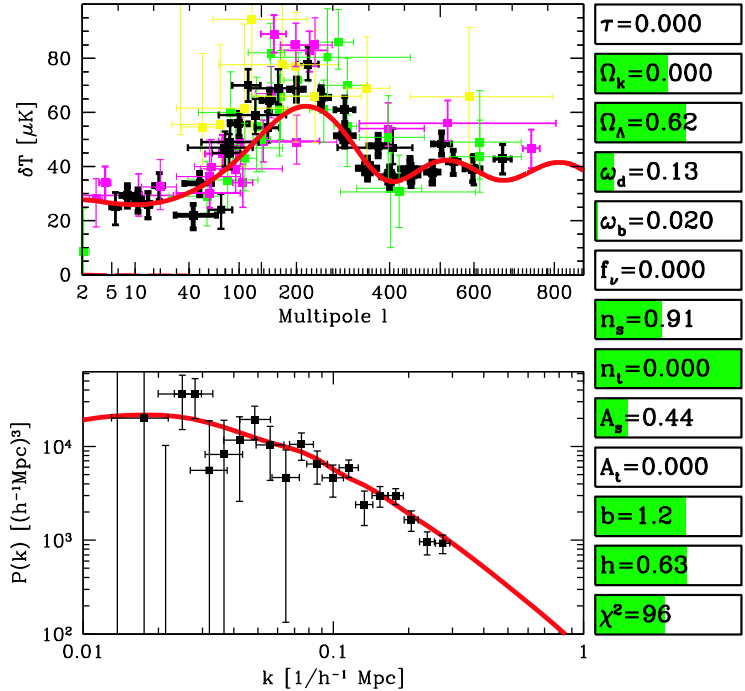}
\end{center}
\caption[]{\footnotesize%
The data used is shown for the CMB (top) and the LSS (bottom)
together with the best fit ``concordance'' model from Table 1.
Animated versions of this figure, where the effect of changing one parameter
at a time can be viewed, are available at 
www.hep.upenn.edu/$\sim$max/concordance.html.
\label{CPfig}}
\end{figure}

\section{INTRODUCTION}

The cosmic microwave background (CMB) is dramatically improving our
knowledge of cosmological parameters
\cite{Lange00,boompa,Bambi00,Bridle00,observables,Jaffe00,Kinney00}.
However, since the CMB still suffers from so-called degeneracies,
where the effect of changing some parameters can be almost
canceled by changing others, even more information can be extracted
if additional information is included in the analysis.
The power spectrum of large scale structure (LSS) in the 
galaxy distribution
is particularly powerful in this regard, since
it is depends on almost all of the parameters that 
affect the CMB, but in different ways since the physics 
involved is different. Since the sources of systematic errors are also
different, a joint CMB+LSS analysis has the additional merit
of allowing a number of consistency checks to be made. 

In this paper, we will perform such a joint CMB+LSS analysis using 
the data shown in \fig{CPfig}.
We include all currently available CMB data \cite{Gawiser00}.
The LSS data is the linear real space power spectrum
of the {\it IRAS} Point Source Catalogue Redshift Survey \cite{Saunders00}
(PSCz)
as measured by \cite{pscz}.
The PSCz survey contains 18{,}351 galaxies covering 84\%
of the sky to a usable depth of about $400 \, h^{-1}\Mpc$.
Earlier CMB+LSS work 
\cite{Efstathiou92,Kofman93,WhiteBaryons,Liddle96,Bunn97,Gawiser98,Webster98,Bond98,CosmicTriangle,Novosyadlyj00,Bridle99}
was recently extended using 
LSS-information
summarized by two parameters \cite{Lange00,Jaffe00}
--- here we treat the LSS data just as the CMB, including the
full power spectrum shape.

\section{METHOD}

\label{MethodSec}

Our goal is to constrain jointly the 11 
cosmological parameters 
\beq{pEq}
\p\equiv(\tau,\Ok,\Ol,\od,\ob,\fn,\ns,\nt,\As,\At,b).
\eeq
These are the reionization optical depth $\tau$, 
the primordial amplitudes $\As$, $\At$ and tilts $\ns$, $\nt$ 
of scalar and tensor fluctuations, 
a bias parameter $b$ defined as the ratio between rms 
galaxy fluctuations and rms matter fluctuations on 
large scales,
and five parameters specifying the cosmic matter budget.
The various contributions $\Omega_i$ to critical density are for
curvature $\Ok$, vacuum energy $\Ol$, cold dark matter $\Oc$, 
hot dark matter (neutrinos) $\On$ and baryons $\Ob$.
The quantities
$\ob\equiv h^2\Ob$ and
$\od\equiv h^2\Od$ correspond to 
the physical densities of baryons
and total (cold + hot) dark matter 
($\Od\equiv\Oc+\On$), and $\fn\equiv\On/\Od$ is the fraction
of the dark matter that is hot.
We assume that the bias $b$ is constant on large scales 
but make no assumptions about its value, 
and therefore marginalize over this parameter
before quoting constraints on the other ten.

Just as in particle physics, it is possible to extend this 
``minimal standard model'' by introducing more physics and more parameters.
We limit our analysis to these 11 since they are all so 
well-motivated theoretically or observationally that it would be 
inappropriate to leave them out or to assume that we know their values
{\it a priori}.

Our method consists of the following steps:
\begin{enumerate}
\item Compute power spectra $C_\l$ and $P(k)$ for a grid
of models in our 11-dimensional parameter space.
\item Compute a likelihood for each model that quantifies how well it fits the
data.
\item Perform 11-dimensional interpolation and marginalize to obtain
constraints on individual parameters and parameter pairs.
\end{enumerate}
To make step 1 feasible in practice, we recently developed a method 
for accelerated power spectrum calculation \cite{concordance} that
speeds up the widely used CMBfast software
\cite{cmbfast} by a factor around $10^3$ without appreciable loss
of accuracy. The details of our calculations and assumptions
are given in the above-mentioned
paper --- the results below summarize the parts of the conclusions
of greatest interest to a particle physics audience.

\section{RESULTS}

\label{ResultsSec}

\subsection{Basic results}

\begin{figure}[tb]
\begin{center}
\includegraphics[width=\bredd\textwidth]{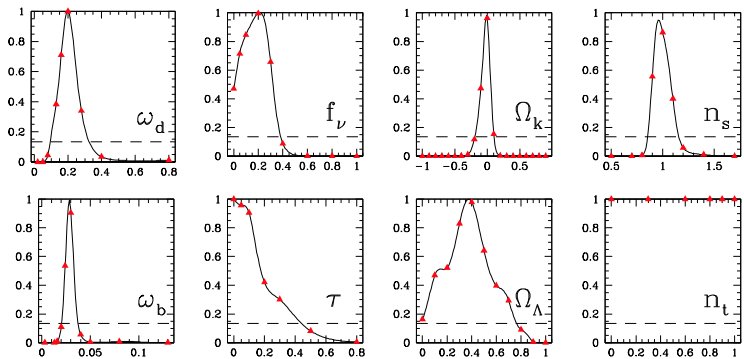}
\end{center}
\caption[]{
Constraints on individual parameters using only CMB and LSS information.
The quoted 95\% confidence limits are where each curve drops below
the dashed line.
}
\label{1DnoFig}
\end{figure}

Our constraints on individual cosmological parameters are 
listed in Table 1 for three cases and plotted in 
\fig{1DnoFig} for them.
All tabulated and plotted bounds are 95\% confidence limits.
The first case uses constraints from CMB alone, which are 
still rather weak because of degeneracy problems.
The second case combines the CMB information with 
the power spectrum measurements from PSCz, and is seen to
give rather interesting constraints on most parameters
except the tensor tilt $\nt$.
The third case adds three assumptions:
that the latest measurements of
the baryon density $\ob=0.019\pm 0.0024$ 
from Big Bang Nucleosynthesis (BBN) are correct
\cite{Burles99},
that the $1\sigma$ constraints on the 
Hubble parameter are $h=0.74\pm 0.08$ 
\cite{Freedman00},
and
that the neutrino contribution is cosmologically negligible.
The neutrino assumption is that there is no strong mass-degeneracy between
the relevant neutrino families, and that the Super-Kamiokande atmospheric
neutrino data therefore sets the scale of the 
neutrino density to be $\on\sim\times 10^{-4}-10^{-3}$ 
\cite{Scholberg99}.
We emphasize that this last assumption 
(that the heaviest neutrino
weighs of order the root of the squared mass difference
$\Delta m^2\sim 0.07\eV^2$)
is merely motivated by Occam's razor,
not by observational evidence --- 
the best current limits on $\fn$ from other astrophysical
observations (see \cite{Croft99}
and references therein) are
still compatible with $\fn\sim 0.2$.
Rather, we have chosen to highlight the consequences
of this prior since, as discussed below, 
it has interesting effects on other parameters.

\begin{table}

\bigskip
\noindent
{\footnotesize
{\bf Table 1} -- Best fit values and 95\% confidence limits on
cosmological parameters.
The ``concordance'' case combines CMB and PSCz information with
a BBN prior $\ob=0.02$, a Hubble prior $h=0.74\pm 0.08$ and 
a prior that $\fn\sim 10^{-3}$. A dash indicates that no meaningful
constraint was obtained.
The redshift space distortion parameter is
$\beta\equiv f(\Om,\Ol)/b$, where $f$ is the linear growth rate.
$\zion$ is the redshift of reionization and $t_0$ is the present age
of the Universe in Gigayears.

\def\fnp{{\it$\sim$0}}
\smallskip

{
\begin{center}
\begin{tabular}{|l|ccc|ccc|ccc|}
\hline
			&\multicolumn{3}{c|}{CMB alone}
			&\multicolumn{3}{c|}{CMB + PSCz}
			&\multicolumn{3}{c|}{Concordance}\\
Quantity		&Min	&Best	&Max	&Min	&Best	&Max	&Min	&Best	&Max\\
\hline
$\tau$			&0.0	&0.0	&$0.32$	&0.0	&0.0	&$.44$	&0.0	&0.0	&$.16$	\\
$\Ok$			&$-$.69	&$-$.34	&0.05	&$-$.19	&$-$.02	&0.10	&$-$.05	&$-$.00	&0.08	\\
$\Ol$			&$.05$	&.43	&$.92$	&$-$	&.38	&0.76	&.49	&.62	&0.74	\\
$h^2\Od$		&0.0	&.10	&$-$	&.10	&.20	&0.32	&.11	&.13	&0.17	\\
$h^2\Ob$		&.024	&.054	&$.103$	&.020	&.028	&.037	&{\it .02}&{\it .02}&{\it .02}\\
$\fn$			&0.0	&.80	&$1.0$	&0.0	&.22	&.38	&\fnp	&\fnp	&\fnp	\\
$\ns$			&.91	&1.43	&$-$	&0.86	&.96	&1.16	&0.84	&.92	&1.01	\\
$\nt$			&$-$	&0.0	&$-$	&$-$	&0.0	&$-$	&$-$	&0.0	&$-$	\\
\hline		
$b$			&$-$	&$-$	&$-$	&.75	&1.26	&1.78	&.87	&1.10	&1.33	\\
$h$			&.18	&.53	&.88	&.33	&.59	&.86	&.58	&.68	&.78	\\
$\beta$			&$-$	&$-$	&$-$	&.37	&.63	&.89	&.36	&.51	&.66	\\
$\zion$			&0	&7	&21	&0	&9	&$26$	&0	&6	&20	\\
$t_0$			&8.4	&15.6	&23.0	&9.6	&13.3	&17.0	&12.1	&13.4	&14.6	\\
\hline		
\end{tabular}
\end{center}
}
}
\end{table}

\subsection{Effects of priors}

The 
joint CMB+PSZc constraints are remarkably robust to prior assumptions.
Imposing priors such as flatness ($\Ok=0$), 
no tensors ($r=0$), no tilt ($\ns=1$), no reionization ($\tau=0$),
and a reasonable Hubble parameter
(we tried both $h=0.74\pm 8$ at 65\% 
and the weaker constraint $50<h<100$ at 95\%),
both alone and in various combinations, 
has little effect.
The fact that the best fit parameter
values are not appreciably altered reflects that these priors
all agree well with what is already borne out by the
CMB+PSCz data: $\Ok\sim r\sim\tau\sim 0$, 
and $\ns\sim 1$.
The fact that these priors do not shrink the error bars
much on other parameters indicates that
PSCz has already broken the main CMB degeneracies.

\begin{figure}[tb]
\begin{center}
\includegraphics[width=\bredd\textwidth]{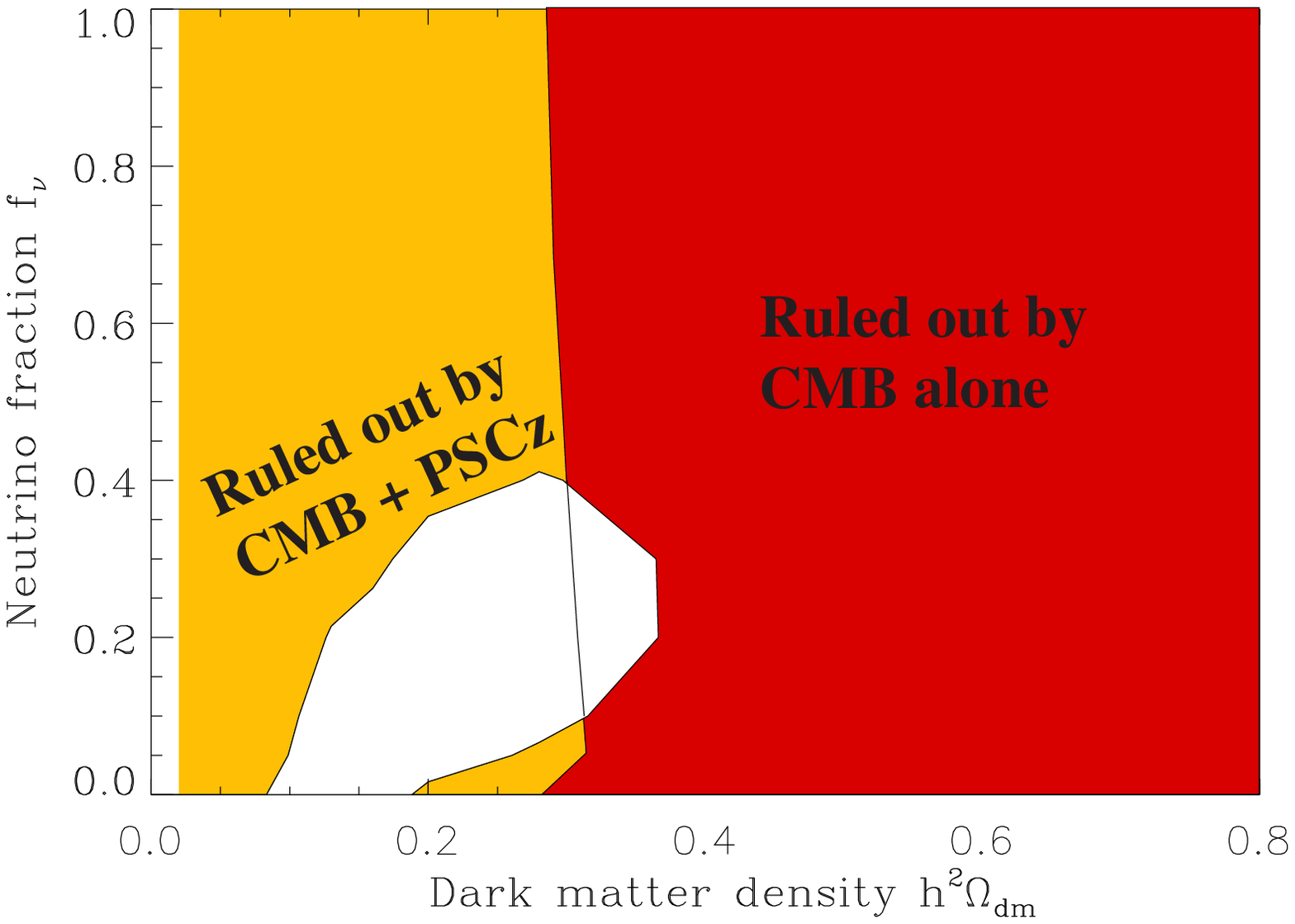}
\end{center}
\caption[]{
Constraints in the $(\od,\fn)$-plane.
The shape of the allowed (white) region explains why the prior 
$\fn=0$ tightens the upper limit on the dark matter density.
The vertical line shows the CMB-only boundary before PSCz is added.
}
\label{odfnFig}
\end{figure}

The nucleosynthesis prior has a greater influence
because it does not agree all that well with what the
CMB+LSS data prefer. 
We found one additional prior that had a non-negligible effect:
that on neutrinos. As illustrated in \fig{odfnFig}, inclusion of 
neutrinos substantially weakens the upper limits on the dark matter
density.
Since the neutrino fraction $\fn$ has only a weak effect on the CMB, 
this effect clearly comes from LSS. 
A larger dark matter density $\od$ pushes matter-radiation
equality back to an earlier time, shifting the corresponding 
turnover in $P(k)$ to the right and thereby increasing the
ratio of small-scale to large-scale power.
Increasing the neutrino fraction counteracts this by suppressing
the small-scale power (without affecting the CMB much), 
thereby weakening the upper limit on $\od$.
Imposing the prior $\fn=0$ alone,
without nucleosynthesis or Hubble priors,
tightens the CMB+PSCz constraint $\od<0.32$ from Table 1 to $\od<0.19$.

\begin{figure}[tb]
\begin{center}
\includegraphics[width=\bredd\textwidth]{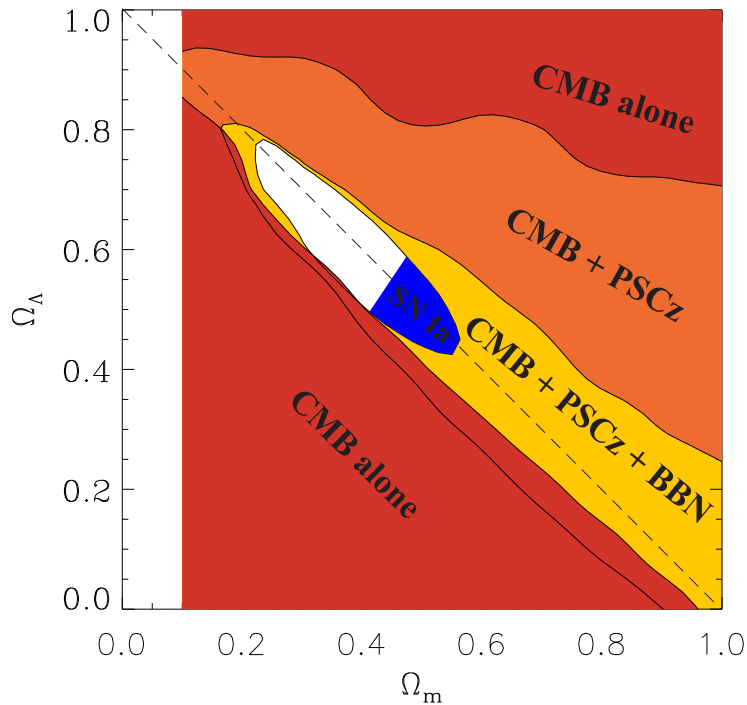}
\end{center}
\caption[]{
Constraints in the $(\Om,\Ol)$-plane.
The shaded regions are ruled out at 95\% confidence by the
information indicated. The allowed (white) region is seen to be centered
around flat models, which fall on the dashed line.
}
\label{OmOlFig}
\end{figure}

\begin{figure}[tb]
\begin{center}
\includegraphics[width=\bredd\textwidth]{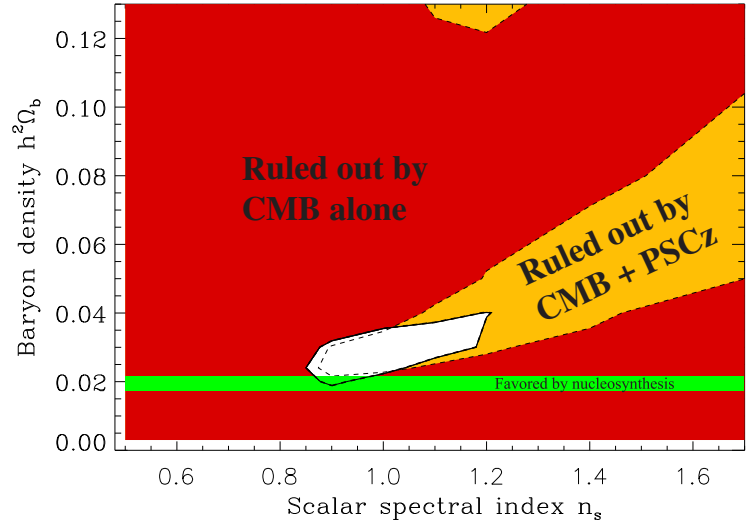}
\end{center}
\caption[]{
Constraints in the $(\ns,\ob)$-plane. Note that PSCz not only shrinks
the allowed region (white), but also pushes it slightly down to the left
(the dashed line indicates the CMB-only boundary).
}
\label{nsobFig}
\end{figure}

\begin{figure}[tb]
\begin{center}
\includegraphics[width=\bredd\textwidth]{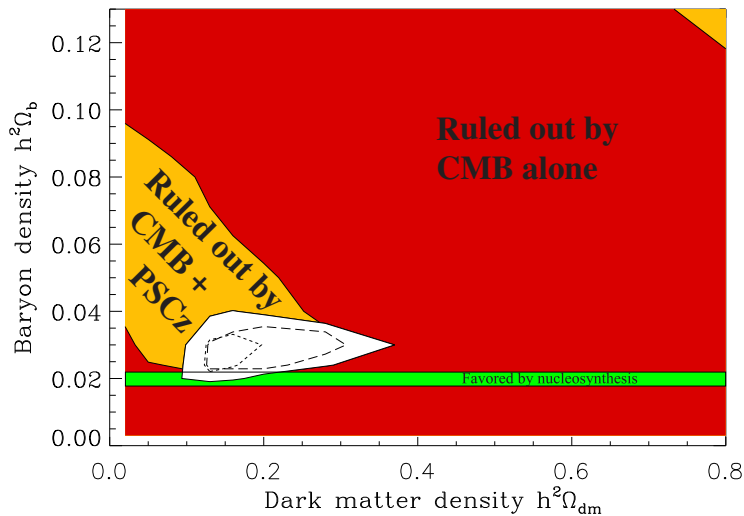}
\end{center}
\caption[]{
Constraints in the $(\od,\ob)$-plane.
As in the previous figure, adding PSCz prohibits 
high baryon solutions and allows slightly 
lower $\ob$-values than CMB alone.
The dashed curve within the allowed (white) region
show the sharper constraint obtained when 
imposing the priors
for a flat, scalar scale-invariant model 
($\Ok=r=0$, $\ns=1$). 
The dotted curve shows the effect of requiring 
negligible neutrino density ($\fn\sim 0$) in addition.
}
\label{odobFig}
\end{figure}

\section{DISCUSSION}

\label{ConclusionsSec}

We have presented joint constraints on 11 cosmological parameters from current 
CMB and galaxy clustering data. 
Perhaps the most interesting results of this paper are the numbers themselves,
listed in the CMB+LSS columns of Table 1, and their striking robustness to 
imposing various priors. A superficial glance at the constraint figures 
might suggest that little has changed since the first analysis  
of Boomerang + Maxima \cite{boompa}, or even since the pre-Boomerang analysis
of \cite{10par}, since the plots look rather similar. 
However, whereas these earlier
papers obtained strong constraints only with various poorly 
justified priors such as no tensors, no tilt or no curvature, 
the joint CMB + LSS data are now powerful enough to 
speak for themselves, without needing any such prior props.

\subsection{Towards a refined concordance model}

It is well-known that different types of measurements can
complement each other by breaking degeneracies. 
However, even more importantly, multiple data sets allow numerous
consistency checks to be made. The present results allow a number of
such tests.

\subsubsection{Baryons}

Perhaps the most obvious one involves the baryon fraction. 
Although there is still some tension between BBN (preferring 
$\ob\sim 0.02$) and CMB+LSS (preferring $\ob\sim 0.03$),
an issue which will undoubtedly be clarified by improved data within 
a year, the most striking point is that the methods 
agree as well as they do. That one 
method involving nuclear physics when the Universe was 
a minute
old and another involving plasma physics more than 
100{,}000 years
later give roughly consistent answers, despite involving
completely different systematics, can hardly be described as anything short
of a triumph for the Big Bang model. 
 
It is noteworthy that our addition of LSS 
information pulls down the baryon value slightly, so that
a BBN-compatible value $\ob=0.02$ is now within the 95\% confidence interval.
Part of the reason that that the CMB alone gave a stronger lower 
limit may be a reflection
of the Bayesian likelihood procedure employed in this and all other recent papers
on the topic: when a large space of high $\ob$-values are allowed, the relative
likelihood for lower values drops.

\subsubsection{Dark energy}

Another important cross-check involves the cosmological constant.
Although the constraint $0.49<\Ol<0.74$ from Table 1 does not involve 
any supernova information, it agrees nicely with the recent
accelerating universe predictions from 
SN 1a \cite{Perlmutter98,Riess98}. This agreement is illustrated in 
\fig{OmOlFig}, which shows the SN 1a constraints from 
\cite{WhiteConcordance} combining the data from 
both teams.
As frequently pointed out, the conclusion $\Om\sim 0.35$ also
agrees well with a number of other observations, \eg, 
the cluster abundance at various redshifts 
and cosmic velocity fields.

\subsubsection{Bias}

A third cross-check is more subtle but equally striking,
involving the bias of the PSCz galaxies --- we can measure
it in two completely independent ways.
One is by comparing the amplitude of the 
CMB and galaxy power spectra, which gives the constraints listed
in Table 1.
The other way is via linear redshift space distortions.
In terms of the redshift space distortion parameter
$\beta\approx\Om^{0.6}/b$, the former method gives
$\beta=0.51\pm 0.08$
and the latter gives $\beta=0.45^{+.14}_{-.12}$ (68\%).
This striking agreement means that a highly non-trivial consistency test
has been passed.

\subsection{Concordance}

In conclusion, the simple ``concordance'' model 
in the last columns of Table 1 (plotted in \fig{CPfig})
is at least marginally consistent with all basic cosmological constraints, 
including CMB, PSCz and nucleosynthesis.
Specifically, as discussed above, our calculations show that 
it 
has passed three non-trivial consistency tests.
Moreover our concordance model is encouragingly robust towards
imposing a score of prior constraints in various combinations.
In summary, the non-baryonic matter that is the topic of this conference
really seems to be out there, and we now have quite strong indications
of what its mean density is:
$2.4\times 10^{-27}\kg/\meter^3$ give or take 20\%.

\bigskip
The authors wish to thank 
Ang\'elica de Oliveira-Costa, Brad Gibson, Wayne Hu and Nikhil Padmanabhan
for useful discussions.
Support for this work was provided by
NSF grant AST00-71213, 
NASA grants NAG5-7128 
and NAG5-9194,
the University of Pennsylvania Research Foundation,
and 
Hubble Fellowship HF-01116.01-98A from 
STScI, operated by AURA, Inc. 
under NASA contract NAS5-26555.

\end{document}